\documentclass[epj, nopacs]{svjour}
\usepackage{graphics}
\usepackage{amsmath}
\usepackage{amssymb}
\usepackage{bm}
\usepackage{slashed}
\begin{document}
\title{Three-dimensional parton distribution functions $g_{1T}$ and $h_{1L}^\perp$ in the polarized proton-antiproton Drell-Yan process}
\author{Jiacai Zhu\inst{1} \and Bo-Qiang Ma\inst{1}$^,$\inst{2}$^,$
\thanks{Corresponding author. Email address: \texttt{mabq@pku.edu.cn}}%
}                     
%
%
\institute{School of
Physics and State Key Laboratory of Nuclear Physics and Technology,
Peking University, Beijing 100871, China \and Center for High Energy
Physics, Peking University, Beijing 100871,
China}
\date{Received: date / Revised version: date}
%
\abstract{
We present predictions of the unweighted and weighted double spin
asymmetries related to  the transversal helicity distribution
$g_{1T}$ and the longitudinal transversity distribution
$h_{1L}^\perp$, two of eight leading-twist transverse momentum
dependent parton distributions (TMDs) or three-dimensional parton
distribution functions (3dPDFs), in the polarized proton-antiproton
Drell-Yan process at typical kinematics on the Facility for
Antiproton and Ion Research (FAIR). We conclude that FAIR is ideal
to access the new 3dPDFs towards a detailed picture of the
nucleon structure.
%
} 
\titlerunning{3dPDFs $g_{1T}$ and $h_{1L}^\perp$ in the polarized proton-antiproton Drell-Yan process}
\authorrunning{Jiacai Zhu and Bo-Qiang Ma}
\maketitle
%
\section{Introduction}
The nucleon spin structure is an active direction under both
theoretical and experimental investigations. A number of new
physical quantities of the nucleon, which can provide a detailed
picture of the nucleon with more detailed information in three
dimensional momentum space~\cite{Barone2002,Barone2010}, have been
introduced. At leading twist, the quark-quark correlation
matrix~\cite{Mulders1996,Boer1998} can be decomposed into eight
transverse momentum dependent parton distributions (TMDs), or we
call them three-dimensional parton distribution functions (3dPDFs).
Besides the three usual parton distribution functions, \textit{i.e.},
the unpolarized one with light-cone longitudinal momentum and
transverse momentum distribution $f_1$, the helicity one with longitudinal
helicity distribution $g_{1L}$, and the transversity one with transversal
spin distributions $h_1$, there are five new ones. Among them the Sivers
distribution $f_{1T}^\perp$ 
and its chiral-odd partner, the
Boer-Mulders distribution $h_1^\perp$, 
are well known for their T-odd property, \textit{i.e.}, they change
sign under naive time reversal. The other three of the new 3dPDFs,
\textit{i.e.}, the pretzelosity distribution $h_{1T}^\perp$, the
transversal helicity distribution $g_{1T}$, and the longitudinal
transversity distribution $h_{1L}^\perp$, are T-even. They can be
measured through the semi-inclusive deep inelastic
scattering~\cite{She2009,Zhu2011}. However, the single spin
asymmetry related to the pretzelosity distribution $h_{1T}^\perp$ in
the semi-inclusive deep inelastic scattering is rather small. For
the chiral-odd distributions $h_{1T}^\perp$ and $h_{1L}^\perp$, they
can be probed through the single spin asymmetries when combined with
another chiral-odd distribution $h_1^\perp$ in the pion-nucleon
Drell-Yan process\cite{Lu2011,Lu2011a}. In Ref.~\cite{Zhu2010}, it
has been shown that the polarized proton-antiproton Drell-Yan
process is ideal to probe the pretzelosity distribution
$h_{1T}^\perp$ and the magnitudes of the related spin asymmetries
are significantly large. Thus it is natural to discuss the spin
asymmetries related to the rest two 3dPDFs, $g_{1T}$ and
$h_{1L}^\perp$, in the polarized proton-antiproton Drell-Yan
process.

Since the proposal on measuring the transversity distributions via the
polarized proton-antiproton Drell-Yan process by the polarized
antiproton experiment (PAX) collaboration~\cite{Barone2005,Efremov2004}, there
have been some other
experiments~\cite{hermes2005,hermes2010,compass2005,compass2010,jlab,Gao2011,Qian2011}
on the measurements of the transversity distributions. Recently, there
has been new technical progress~\cite{Lenisa2010} towards the goal
for a proton-antiproton collider with both beams
polarized~\cite{Rathmann2005}, and such plan has the potential to be
realized at FAIR (Facility for Antiproton and Ion Research) in GSI
Helmholtzzentrum f\"ur Schwerionenforschung. The expected antiproton beam polarizations might be $0.15\sim0.20$ (spin filtering with transverse target orientation) or $0.35\sim0.40$ (longitudinal)~\cite{Barschel2009,Dmitriev2010}. Thus the three new
3dPDFs,  $h_{1T}^\perp$, $g_{1T}$, and $h_{1L}^\perp$, with
important information on the quark spin and orbital correlation of
the nucleon, could be measured on FAIR. Our calculation below shows
that some of the double spin asymmetries related to $g_{1T}$ and
$h_{1L}^\perp$ in the polarized proton-antiproton Drell-Yan process
are large, and thus FAIR is an ideal facility to access the new
3dPDFs for revealing more information towards a detailed picture
of the nucleon.

\section{T-even 3dPDFs in the light-cone quark-diquark model}
In the light-cone quark-diquark model~\cite{Ma1996}, the
Melosh-Wigner rotation plays an important role to understand the
proton spin puzzle~\cite{Ma1991,Ma1993} due to the relativistic
effect of quark transversal motions. The T-even 3dPDFs have been
calculated~\cite{She2009,Zhu2011,Ma1996,Schmidt1997,Ma1998,Ma2000}:
\begin{align}
f_1^{(uv)}(x, k_T^2) &= \frac{1}{16\pi^3}   (\frac{1}{3} \sin^2\theta_0 \varphi_V^2 + \cos^2\theta_0 \varphi_S^2),\nonumber\\
f_1^{(dv)}(x, k_T^2) &= \frac{1}{8\pi^3}  \frac{1}{3} \sin^2\theta_0 \varphi_V^2, \label{eq:unpol}
\end{align}
and
\begin{align}
j^{(uv)}(x, k_T^2)  =& -\frac{1}{16\pi^3} \times (\frac{1}{9} \sin^2\theta_0 \varphi_V^2 W_V^j - \cos^2\theta_0 \varphi_S^2 W_S^j),\nonumber\\
j^{(dv)}(x, k_T^2)  =& -\frac{1}{8\pi^3} \times \frac{1}{9} \sin^2\theta_0 \varphi_V^2 W_V^j, \label{eq:j1}
\end{align}
with the notation
\[
j = g_{1L},\, g_{1T},\, h_1,\, h_{1T}^\perp,\, h_{1L}^\perp,
\]
and the superscripts ``$uv$" and ``$dv$" stand for the valence up and down quark distributions respectively.
$\varphi_D (D = V, S)$ is the wave function in the momentum space for
the quark-diquark, and for which we can use the Brodsky-Huang-Lepage
(BHL) prescription~\cite{Brodsky1982,Huang1994}:
\begin{equation}
\varphi_D(x, k_T^2)=A_D\exp\Big\{-\frac{1}{8\alpha_D^2}\big[\frac{m_q^2 + k_T^2}{x}+\frac{m_D^2 + k_T^2}{1-x}\big]\Big\}.
\end{equation}
The parameters $\alpha_D = 0.33 ~ \mathrm{GeV}$ (which is the same for $D = V, S$), the quark mass $m_q = 0.33 ~ \mathrm{GeV}$, the diquark mass
$m_S = 0.60 ~ \mathrm{GeV}$, $m_V = 0.80 ~ \mathrm{GeV}$, and $\theta_0=\pi/4$ are adopted for numerical calculation.
$\theta_0$ is the mixing angle that breaks the SU(6) symmetry when $\theta_0 \neq \pi/4$.
The Melosh-Wigner rotation factors $W_D$ ($D = V, S$) are
\begin{equation}
W_D^{g_{1L}}(x, k_T^2) = \frac{(x\mathcal{M}_D + m_q)^2 - k_T^2}{(x\mathcal{M}_D + m_q)^2 + k_T^2},
\end{equation}
\begin{equation}
W_D^{g_{1T}}(x, k_T^2) = \frac{2M_N(x\mathcal{M}_D + m_q)}{(x\mathcal{M}_D + m_q)^2 + k_T^2},
\end{equation}
\begin{equation}
W_D^{h_1}(x, k_T^2) = \frac{(x\mathcal{M}_D + m_q)^2}{(x\mathcal{M}_D + m_q)^2 + k_T^2},
\end{equation}
\begin{equation}
W_D^{h_{1T}^\perp}(x, k_T^2) = -\frac{2M_N^2}{(x\mathcal{M}_D + m_q)^2 + k_T^2},
\end{equation}
\begin{equation}
W_D^{h_{1L}^\perp}(x, k_T^2) = -\frac{2M_N(x\mathcal{M}_D + m_q)}{(x\mathcal{M}_D + m_q)^2 + k_T^2},
\end{equation}
where
\begin{equation}
\mathcal{M}_D = \sqrt{\frac{m_q^2 + k_T^2}{x} + \frac{m_D^2 + k_T^2}{1- x}}.
\end{equation}


Using Eqs.~(\ref{eq:unpol}) and (\ref{eq:j1}), the polarized distributions can be given by the unpolarized distributions as
\begin{align}
j^{(uv)}(x, k_T^2) =& \big[f_1^{(uv)}(x, k_T^2) - \frac{1}{2} f_1^{(dv)}(x, k_T^2)\big] W_S^j(x, k_T^2) \nonumber\\
& -\frac{1}{6} f_1^{(dv)}(x, k_T^2) W_V^j(x, k_T^2),\nonumber\\
j^{(dv)}(x, k_T^2) =& -\frac{1}{3} f_1^{(dv)}(x, k_T^2) W_V^j(x, k_T^2).
\label{eq:j2}
\end{align}

\section{The $g_{1T}$ and $h_{1L}^\perp$ related asymmetries in the polarized proton-antiproton Drell-Yan process}
In the polarized proton-antiproton Drell-Yan process, the cross section is~\cite{Arnold2009}
\begin{align}
&\frac{d\sigma}{dx_a ~ dx_b ~ d\bm{q}_T ~ d\Omega}\nonumber\\
=&\frac{\alpha^2_{em}}{4 Q^2} \Big\{ (1 + \cos^2\theta)F_{UU}^1 + S_{aL} S_{bL} \sin^2\theta\cos2\phi F_{LL}^{\cos2\phi}  \nonumber\\
& ~~~~~~+ \lvert\bm{S}_{aT}\rvert S_{bL} (1+\cos^2\theta) \cos\phi_a F_{TL}^{\cos\phi_a}\nonumber\\
 & ~~~~~~+ S_{aL} \lvert\bm{S}_{bT}\rvert \sin^2\theta\big[\cos(2\phi + \phi_b)F_{LT}^{\cos(2\phi + \phi_b)} \nonumber\\
&~~~~~~~~~~~~+ \cos(2\phi - \phi_b)F_{LT}^{\cos(2\phi - \phi_b)}\big]\nonumber\\
 & ~~~~~~+ \lvert\bm{S}_{aT}\rvert \lvert \bm{S}_{bT} \rvert (1 + \cos^2\theta)  \big[( \cos(\phi_a + \phi_b) F_{TT}^{\cos(\phi_a + \phi_b)} \nonumber\\
&~~~~~~~~~~~~ + \cos(\phi_a - \phi_b) F_{TT}^{ \cos(\phi_a - \phi_b)} ) \big] + \cdots \Big\}.
\end{align}
The subscripts $a$ and $b$ stand for the incoming hadrons in the Drell-Yan
process, and $\phi_a$ and $\phi_b$ are the angles of $\bm{S}_{aT}$
and $\bm{S}_{bT}$ respectively. Other terms will not contribute in our analysis below and we should note that although some of these terms involve $g_{1T}$ or $h_{1L}$, they can be obtained from the terms above by switching the subscript labels $a$ and $b$. The structure functions are
\begin{equation}
F_{UU}^{1} = \mathcal{C} [f_1\bar{f}_1],
\end{equation}
\begin{equation}
F_{LL}^{\cos2\phi} = \mathcal{C} \Big[\frac{2(\bm{h}\cdot\bm{k}_{aT})(\bm{h}\cdot\bm{k}_{bT}) - \bm{k}_{aT}\cdot\bm{k}_{bT}}{M_N^2} h_{1L}^\perp \bar{h}_{1L}^\perp\Big],
\end{equation}
\begin{equation}
F_{TL}^{\cos\phi_a} = \mathcal{C} \Big[-\frac{\bm{h}\cdot\bm{k}_{aT}}{M_N} g_{1T}\bar{g}_{1L}\Big],
\end{equation}
\begin{equation}
F_{LT}^{\cos(2\phi - \phi_b)} = \mathcal{C} \Big[\frac{\bm{h}\cdot\bm{k}_{aT}}{M_N} h_{1L}^\perp \bar{h}_{1}\Big],
\end{equation}
\begin{equation}
\begin{split}
&F_{LT}^{\cos(2\phi + \phi_b)} \\
=& \mathcal{C} \Big[\frac{2(\bm{h}\cdot\bm{k}_{bT})[2(\bm{h}\cdot\bm{k}_{aT})(\bm{h}\cdot\bm{k}_{bT}) - \bm{k}_{aT}\cdot\bm{k}_{bT}] - k_{bT}^2(\bm{h}\cdot\bm{k}_{aT})}{2M_N^3} \\
&~~~~~~ h_{1L}^\perp \bar{h}_{1T}^\perp\Big],
\end{split}
\end{equation}
\begin{equation}
F_{TT}^{\cos(\phi_a + \phi_b)} = \mathcal{C} \Big[\frac{2(\bm{h}\cdot\bm{k}_{aT})(\bm{h}\cdot\bm{k}_{bT}) - \bm{k}_{aT}\cdot\bm{k}_{bT}}{2M_N^2} (f_{1T}^\perp \bar{f}_{1T}^\perp - g_{1T}\bar{g}_{1T})\Big],
\end{equation}
\begin{equation}
F_{TT}^{\cos(\phi_a - \phi_b)} = \mathcal{C}
\Big[-\frac{\bm{k}_{aT}\cdot\bm{k}_{bT}}{2M_N^2} (f_{1T}^\perp
\bar{f}_{1T}^\perp + g_{1T}\bar{g}_{1T})\Big],
\end{equation}
where we use the shorthand notation
\begin{equation}
\begin{split}
&\mathcal{C}\big[w(\bm{k}_{aT},\bm{k}_{bT})f_a\bar{f}_b\big] \\
=& \frac{1}{N_c} \sum_q e_q^2 \int d \bm{k}_{aT} ~ d \bm{k}_{bT} ~ \delta^{(2)}\left(\bm{q}_T - \bm{k}_{aT} - \bm{k}_{bT} \right) \\
 &\times w(\bm{k}_{aT},\bm{k}_{bT}) f^q_a( x_a ,k_{aT}^2) f^{\bar{q}}_b( x_b , k_{bT}^2),
\end{split}
\end{equation}
where both quarks and antiquarks of all flavors are taken into
account during the summation over $q$. The unit vector is defined as
$\bm{h} \equiv \bm{q}_T/q_T$.

\begin{figure*}
\begin{center}
\resizebox{0.9\textwidth}{!}{%
  \includegraphics{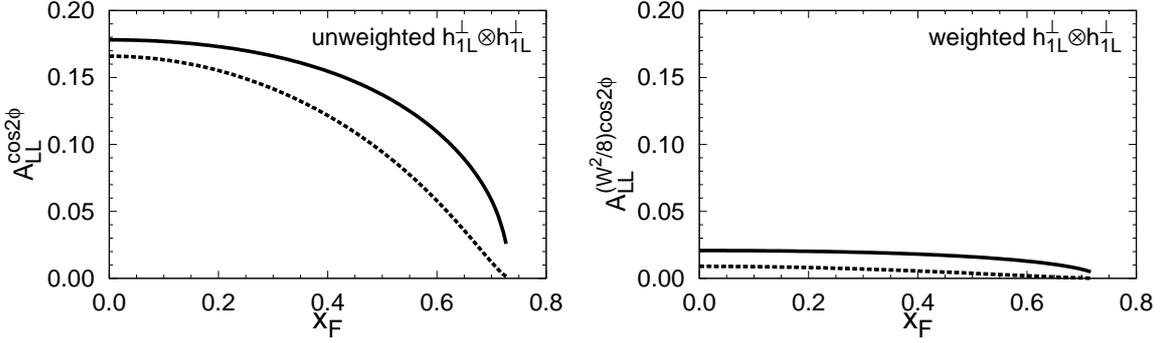}
}
\caption{\label{fig:h_h}The double spin
asymmetries related to $h_{1L}^\perp$ shown in Eqs.~(\ref{eq:asy_h_h}) and (\ref{eq:w_asy_h_h}) as functions of $x_F$ for $s = 45 ~ \mathrm{GeV}^2$ and
$Q^2 = 12 ~ \mathrm{GeV}^2$. The dashed curves correspond to approach 1,
while the solid curves correspond to approach 2.}
\end{center}
\end{figure*}

\begin{figure}
\begin{center}
\resizebox{0.45\textwidth}{!}{%
  \includegraphics{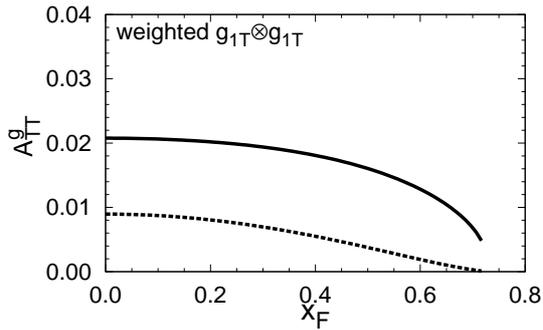}
}
\caption{\label{fig:g_g}The double spin
asymmetry related to $g_{1T}$ shown in Eq.~(\ref{eq:w_asy_g_g}) as a function of $x_F$ for $s = 45 ~ \mathrm{GeV}^2$ and
$Q^2 = 12 ~ \mathrm{GeV}^2$. The dashed curve corresponds to approach 1,
while the solid curve corresponds to approach 2.}
\end{center}
\end{figure}

Considering the charge conjugation invariance
\begin{align}
f_{\bar{p}}^q(x, k_T^2) & = f_p^{\bar{q}}(x, k_T^2), \\
f_{\bar{p}}^{\bar{q}}(x, k_T^2) & = f_p^q(x, k_T^2),
\end{align}
with $p$ for proton and $\bar{p}$ for antiproton, and using the method introduced in Refs.~\cite{Kotzinian1997,Lu2007}, we get
\begin{equation}
\int d\bm{q}_T F_{UU}^{1} = \frac{1}{N_c}\sum_{q} e_q^2 f_1^q(x_a) f_1^q(x_b) ,
\end{equation}
\begin{equation}
\int d\bm{q}_T  \frac{q_T^2}{8M_N^2} F_{LL}^{\cos2\phi} = \frac{1}{N_c}\sum_{q} e_q^2 h_{1L}^{\perp(1)q}(x_a) h_{1L}^{\perp(1)q}(x_b),
\end{equation}
\begin{equation}
\int d\bm{q}_T  \frac{q_T}{2M_N} F_{TL}^{\cos\phi_a} = -\frac{1}{N_c}\sum_{q} e_q^2 g_{1T}^{(1)q}(x_a) g_{1L}^q(x_b),
\end{equation}
\begin{equation}
\int d\bm{q}_T \frac{q_T}{2M_N} F_{LT}^{\cos(2\phi - \phi_b)} = \frac{1}{N_c} \sum_{q} e_q^2 h_{1L}^{\perp(1)q}(x_a) h_1^q(x_b),
\end{equation}
\begin{equation}
\int d\bm{q}_T \frac{q_T^3}{12M_N^3} F_{LT}^{\cos(2\phi + \phi_b)} \\
= \frac{1}{N_c}\sum_{q} e_q^2 h_{1L}^{\perp(1)q}(x_a) h_{1T}^{\perp(2)q}(x_b),
\end{equation}
\begin{equation}
\begin{split}
&\int d\bm{q}_T \frac{q_T^2}{4M_N^2} F_{TT}^{\cos(\phi_a + \phi_b)}  \\
=&\frac{1}{N_c} \sum_{q} e_q^2 \big[f_{1T}^{\perp(1)q}(x_a) f_{1T}^{\perp(1)q}(x_b) - g_{1T}^{(1)q}(x_a) g_{1T}^{(1)q}(x_b)\big],
\end{split}
\end{equation}
\begin{equation}
\begin{split}
&\int d\bm{q}_T \frac{q_T^2}{2M_N^2} F_{TT}^{\cos(\phi_a - \phi_b)} \\
=& - \frac{1}{N_c}\sum_{q} e_q^2 \big[f_{1T}^{\perp(1)q}(x_a) f_{1T}^{\perp(1)q}(x_b) + g_{1T}^{(1)q}(x_a) g_{1T}^{(1)q}(x_b)\big],
\end{split}
\end{equation}
with
\begin{equation}
j^{(n)}(x) \equiv \int d\bm{k}_T \bigg(\frac{k_T^2}{2M_N^2}\bigg)^n j(x, k_T^2),
\end{equation}
for 3dPDF $j$.
\begin{figure*}
\begin{center}
\resizebox{0.9\textwidth}{!}{%
  \includegraphics{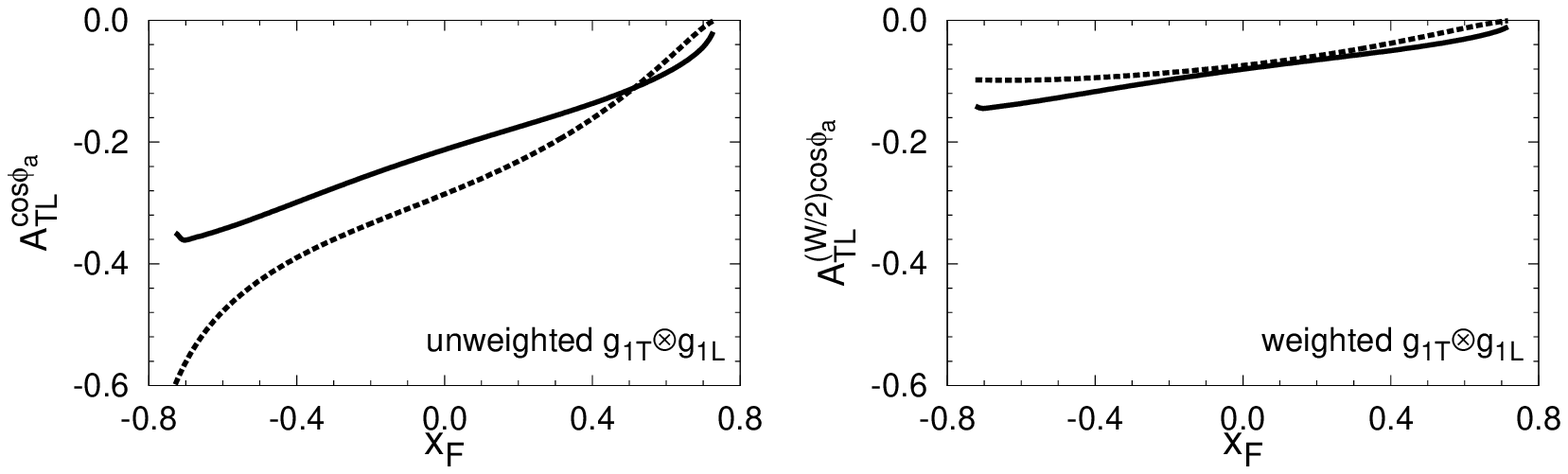}
}
\caption{\label{fig:g_helic}The double spin
asymmetries related to $g_{1T}$ and $g_{1L}$ shown in Eqs.~(\ref{eq:asy_g_helic}) and (\ref{eq:w_asy_g_helic}) as functions of $x_F$ for $s = 45 ~ \mathrm{GeV}^2$ and
$Q^2 = 12 ~ \mathrm{GeV}^2$. The dashed curves correspond to approach 1,
while the solid curves correspond to approach 2.}
\end{center}
\end{figure*}
The unweighted double spin asymmetries are ($Q^2$ is
fixed)
\begin{align}
A_{LL}^{\cos2\phi}(x_F) &= \frac{\int d\bm{q}_T F_{LL}^{\cos2\phi}(x_a, x_b, \bm{q}_T)}{\int d\bm{q}_T F_{UU}^{1}(x_a, x_b, \bm{q}_T)}\nonumber\\
&= \frac{N_c\int d\bm{q}_T F_{LL}^{\cos2\phi}(x_a, x_b, \bm{q}_T)}{\sum_{q} e_q^2 f_1^q(x_a) f_1^q(x_b)} \label{eq:asy_h_h},
\end{align}
\begin{align}
A_{TL}^{\cos\phi_a}(x_F) &= \frac{\int d\bm{q}_T F_{TL}^{\cos\phi_a}(x_a, x_b, \bm{q}_T)}{\int d\bm{q}_T F_{UU}^{1}(x_a, x_b, \bm{q}_T)}\nonumber\\
&= \frac{N_c\int d\bm{q}_T F_{TL}^{\cos\phi_a}(x_a, x_b, \bm{q}_T)}{\sum_{q} e_q^2 f_1^q(x_a) f_1^q(x_b)} \label{eq:asy_g_helic},
\end{align}
\begin{align}
A_{LT}^{\cos(2\phi - \phi_b)}(x_F) &= \frac{\int \bm{q}_T F_{LT}^{\cos(2\phi - \phi_b)}(x_a, x_b, \bm{q}_T)}{ \int \bm{q}_T F_{UU}^{1}(x_a, x_b, \bm{q}_T)}\nonumber\\
&=\frac{N_c\int \bm{q}_T F_{LT}^{\cos(2\phi - \phi_b)}(x_a, x_b, \bm{q}_T)}{\sum_{q} e_q^2 f_1^q(x_a) f_1^q(x_b)}\label{eq:asy_h_trans},
\end{align}
\begin{align}
A_{LT}^{\cos(2\phi + \phi_b)}(x_F) &= \frac{\int\bm{q}_T F_{LT}^{\cos(2\phi + \phi_b)}(x_a, x_b, \bm{q}_T)}{\int \bm{q}_T F_{UU}^{1}(x_a, x_b, \bm{q}_T)}\nonumber\\
&=\frac{N_c\int\bm{q}_T F_{LT}^{\cos(2\phi + \phi_b)}(x_a, x_b, \bm{q}_T)}{\sum_{q} e_q^2 f_1^q(x_a) f_1^q(x_b)}\label{eq:asy_h_pretz}.
\end{align}
The weighted double spin asymmetries are ($Q^2$ is fixed)
\begin{align}
A_{LL}^{\frac{W^2}{8}\cos2\phi}(x_F) &=\frac{\int d\bm{q}_T \frac{W^2}{8} F_{LL}^{\cos2\phi}(x_a, x_b, \bm{q}_T)}{\int d\bm{q}_T F_{UU}^{1}(x_a, x_b, \bm{q}_T)}\nonumber\\
&= \frac{\sum_{q} e_q^2 h_{1L}^{\perp(1)q}(x_a) h_{1L}^{\perp(1)q}(x_b)}{ \sum_{q} e_q^2 f_1^q(x_a) f_1^q(x_b) }\label{eq:w_asy_h_h},
\end{align}
\begin{align}
A_{TL}^{\frac{W}{2}\cos\phi_a}(x_F) &=\frac{\int d\bm{q}_T \frac{W}{2} F_{TL}^{\cos\phi_a}(x_a, x_b, \bm{q}_T)}{\int d\bm{q}_T F_{UU}^{1}(x_a, x_b, \bm{q}_T)}\nonumber\\
&= -\frac{\sum_{q} e_q^2 g_{1T}^{(1)q}(x_a) g_{1L}^q(x_b)}{ \sum_{q} e_q^2 f_1^q(x_a) f_1^q(x_b) }\label{eq:w_asy_g_helic},
\end{align}
\begin{align}
A_{LT}^{\frac{W}{2}\cos(2\phi - \phi_b)}(x_F) &=\frac{\int \bm{q}_T \frac{W}{2} F_{LT}^{\cos(2\phi - \phi_b)}(x_a, x_b, \bm{q}_T)}{ \int \bm{q}_T F_{UU}^{1}(x_a, x_b, \bm{q}_T)}\nonumber\\
&= \frac{\sum_{q} e_q^2 h_{1L}^{\perp(1)q}(x_a) h_1^q(x_b)}{ \sum_{q} e_q^2 \big[f_1^q(x_a) f_1^q(x_b)}\label{eq:w_asy_h_trans},
\end{align}
\begin{align}
A_{LT}^{\frac{W^3}{12}\cos(2\phi + \phi_b)}(x_F) &=\frac{\int\bm{q}_T \frac{W^3}{12} F_{LT}^{\cos(2\phi + \phi_b)}(x_a, x_b, \bm{q}_T)}{\int \bm{q}_T F_{UU}^{1}(x_a, x_b, \bm{q}_T)}\nonumber\\
&= \frac{\sum_{q} e_q^2 h_{1L}^{\perp(1)q}(x_a) h_{1T}^{\perp(2)q}(x_b) }{ \sum_{q} e_q^2 f_1^q(x_a) f_1^q(x_b) }\label{eq:w_asy_h_pretz},
\end{align}
\begin{align}
&A_{TT}^{g}(x_F) \nonumber\\
=&-\frac{\int\bm{q}_T \frac{W^2}{8}F_{TT}^{\cos(\phi_a + \phi_b)}(x_a, x_b, \bm{q}_T)}{\int \bm{q}_T F_{UU}^{1}(x_a, x_b, \bm{q}_T)} \nonumber\\
&~~~~~~ - \frac{\int\bm{q}_T \frac{W^2}{4} F_{TT}^{\cos(\phi_a - \phi_b)}(x_a, x_b, \bm{q}_T)}{\int \bm{q}_T F_{UU}^{1}(x_a, x_b, \bm{q}_T)}\nonumber\\
 =& \frac{\sum_{q} e_q^2 g_{1T}^{(1)q}(x_a) g_{1T}^{(1)q}(x_b)}{ \sum_{q} e_q^2 f_1^q(x_a) f_1^q(x_b) }\label{eq:w_asy_g_g},
\end{align}
where $W=q_T/M_N$. In Eq.~(\ref{eq:w_asy_g_g}),  the weight function is complicated and we use ``$g$" to denote it in Fig.~\ref{fig:g_g}. $x_a$ and $x_b$ are given by
\begin{equation}
x_F = x_a - x_b,~x_ax_b = \frac{Q^2}{s}.
\end{equation}
In Ref.~\cite{Zhu2010}, the $h_{1T}^\perp$ related double spin asymmetries in the polarized proton-antiproton Drell-Yan process have been calculated at typical kinematics on FAIR~\cite{Barone2005}. In this paper, we focus on predictions of the unweighted and weighted double spin asymmetries related to $g_{1T}$ and $h_{1L}^\perp$ in the same process. We present numerical calculations in two different approaches as described in Ref.~\cite{Zhu2010}:
\begin{itemize}
\item \textit{Approach 1}. We use Eqs.~(\ref{eq:unpol}) and (\ref{eq:j1}) directly to calculate and only sum over the valence quark distributions.

\item \textit{Approach 2}. For the unpolarized quark and antiquark distributions, we use the CTEQ6L parametrization~\cite{Pumplin2002}, and
adopt a Gaussian form factor for the transverse
momentum dependence which has been adopted in
many phenomenological analysis~\cite{Anselmino2009}:
\begin{equation}
f_1(x, k_T^2) = f_1(x) \frac{\exp(-k_T^2 / k_{un}^2)}{\pi k_{un}^2}
\end{equation}
with $k_{un}^2 = 0.25~\mathrm{GeV}^2$. For the polarized distributions, we keep
the relations in Eqs.~(\ref{eq:j2}) which we get in the light-cone quark-diquark
model so that the Melosh-Wigner rotation factors remain as the relativistic
effect of quark transversal motions.
\end{itemize}
The kinematics on FAIR are chosen as $s = 45~\mathrm{GeV}^2$ and
$Q^2 = 12~\mathrm{GeV}^2$~\cite{Barone2005}. The magnitudes of the
helicity and transversity distributions in approach 2 are comparable
with the global analysis
results~\cite{Hirai2006,Florian2008,Florian2009,Leader2010,Blumlein2010}
for helicity and~\cite{Anselmino2009} for transversity at the middle
$x$ region. Besides, the quark-diquark model gives a good
description of the nucleon form-factors~\cite{Ma2002a,Ma2002b}. The
quark-diquark model realized in approach 2 also provides reasonable
descriptions of many experiments related to helicity
distributions~\cite{Chen2005}, transversity
distributions~\cite{Huang2007,Qian2011}, together with some new
3dPDFs~\cite{Huang2011}. The effect of the CTEQ6L parametrization,
which has been well verified and constrained by many experiments
concerning the unpolarized quark distributions, has been taken into
account in approach 2. Thus approach 2 might give more reasonable
predictions for future experiments.

\begin{figure*}
\begin{center}
\resizebox{0.9\textwidth}{!}{%
  \includegraphics{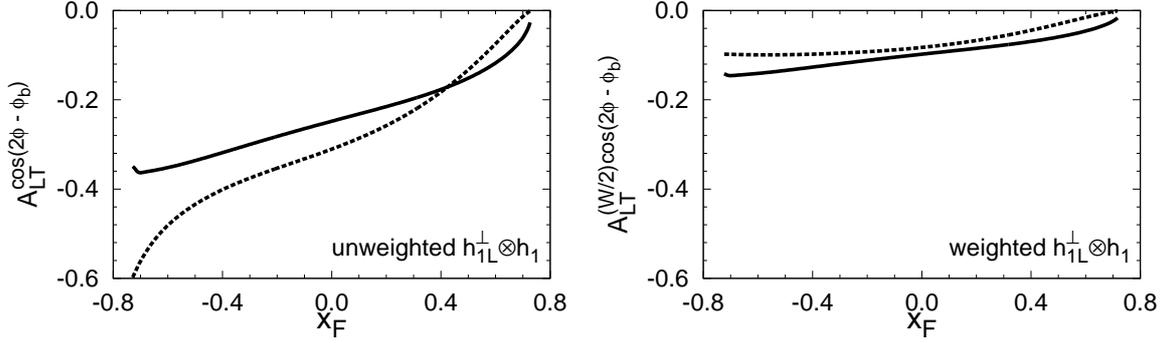}
}
\caption{\label{fig:h_trans}The double spin asymmetries related to $h_{1L}^\perp$  and $h_1$ shown in Eqs.~(\ref{eq:asy_h_trans}) and (\ref{eq:w_asy_h_trans}) as functions of $x_F$ for $s = 45 ~ \mathrm{GeV}^2$ and
$Q^2 = 12 ~ \mathrm{GeV}^2$. The dashed curves correspond to approach 1,
while the solid curves correspond to approach 2.}
\end{center}
\end{figure*}

\begin{figure*}
\begin{center}
\resizebox{0.9\textwidth}{!}{%
  \includegraphics{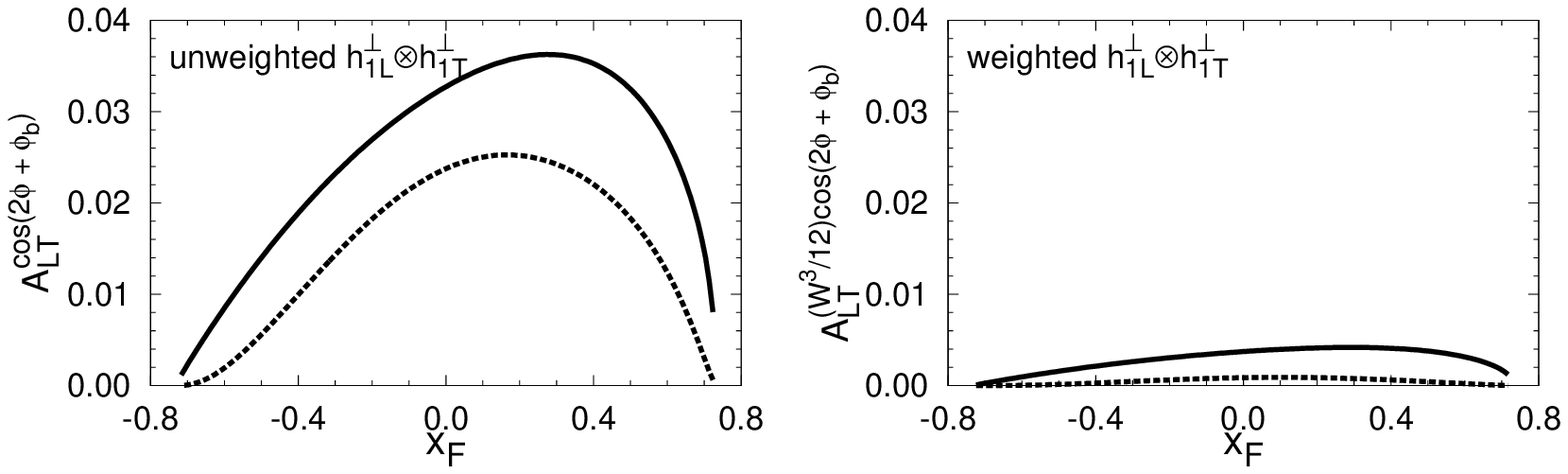}
}
\caption{\label{fig:h_pretz}The double spin asymmetries related to $h_{1L}^\perp$ and $h_{1T}^\perp$ shown in Eqs.~(\ref{eq:asy_h_pretz}) and (\ref{eq:w_asy_h_pretz}) as functions of $x_F$ for $s = 45 ~ \mathrm{GeV}^2$ and
$Q^2 = 12 ~ \mathrm{GeV}^2$. The dashed curves correspond to approach 1,
while the solid curves correspond to approach 2.}
\end{center}
\end{figure*}

We plot the unweighted and weighted asymmetries related to $g_{1T}$ and
$h_{1L}^\perp$ as show in Eqs. (\ref{eq:asy_h_h}), (\ref{eq:asy_g_helic}), (\ref{eq:asy_h_trans}),
(\ref{eq:asy_h_pretz}), (\ref{eq:w_asy_h_h}), (\ref{eq:w_asy_g_helic}), (\ref{eq:w_asy_h_trans}),
(\ref{eq:w_asy_h_pretz}), and (\ref{eq:w_asy_g_g}). The results are shown in
Figs.~\ref{fig:h_h}, \ref{fig:g_helic}, \ref{fig:h_trans},
\ref{fig:h_pretz}, and \ref{fig:g_g}. In the light-cone quark-diquark model, the
ratios of $h_{1L}^{\perp(1/2)}(x)/f_1(x)$ as shown in Ref.~\cite{Zhu2011} are larger
compared with those of $h_{1T}^{\perp(1)}(x)/f_1(x)$ as shown in
Ref.~\cite{She2009}, so the unweighted double spin asymmetry related to
$h_{1L}^\perp$ and $h_1$ in Fig.~\ref{fig:h_trans} is correspondingly larger
compared with the unweighted double spin asymmetry related to $h_{1T}^\perp$
and $h_1$ that we obtained in Ref.~\cite{Zhu2010}. The unweighted asymmetries
are larger compared with the corresponding weighted ones. In
Figs.~\ref{fig:h_h}, \ref{fig:g_helic} and \ref{fig:h_trans}, the unweighted
asymmetries are significantly large. Besides, the unique feature of the unweighted
double spin asymmetry $A_{LL}^{\cos2\phi}$ in Eq.~(\ref{eq:asy_h_h}) as shown in
Fig.~\ref{fig:h_h} is that only the new 3dPDF $h_{1L}^\perp$ is involved. Thus it is
ideal to measure the
new 3dPDFs $g_{1T}$ and $h_{1L}^\perp$ through the double spin asymmetries
in the polarized proton-antiproton Drell-Yan process. Moreover, the unweighted
double spin asymmetry related to $h_{1L}^\perp$ and $h_{1T}^\perp$ in
Fig.~\ref{fig:h_pretz} is about several percent.
This provides us a way to measure the pretzelosity
distributions $h_{1T}^\perp$ in the proton-antiproton Drell-Yan process with one
nucleon longitudinal polarized and another one transversal polarized, which is
different from the unweighted double spin asymmetry related to $h_{1T}^\perp$
and $h_1$ in the proton-antiproton Drell-Yan process with
both nucleons transversal polarized in Ref.~\cite{Zhu2010}.

\section{Summary}
$g_{1T}$ and $h_{1L}^\perp$, \textit{i.e.}, the transversal helicity
and the longitudinal transversity, are two of the eight
leading-twist 3-dimensional parton distributions (3dPDFs). We
present predictions of the unweighted and weighted double spin
asymmetries related to them in the polarized proton-antiproton
Drell-Yan process at typical kinematics on FAIR respectively. We
conclude that the Facility for Antiproton and Ion Research (FAIR) is
ideal to access the new 3dPDFs towards a detailed picture of the
nucleon structure.

\begin{acknowledgement}
This work is supported by National Natural Science Foundation of
China (Grants No.~11021092, No.~10975003, No.~11035003, and
No.~11120101004).
\end{acknowledgement}

%
%


\end{document}